\documentclass[prb,twocolumn,showpacs]{revtex4-1}
\usepackage{graphicx}
\usepackage{amsmath}
\usepackage{amssymb}
\usepackage{epstopdf}
\usepackage{slashed}

\begin{document}
\title{Flat band electrons and interactions in rhombohedral trilayer graphene}
\author{Hao Wang,$^1$ Jin-Hua Gao,$^{2,1}$ and Fu-Chun Zhang$^{1,3}$}
\affiliation{$^1$Department of Physics, The University of Hong Kong, Hong Kong SAR, China\\
$^2$Department of Physics, Huazhong University of Science and Technology, Wuhan, China\\
$^3$Department of Physics, Zhejiang University, Hangzhou, 310027,
China}

\begin{abstract}
Multilayer graphene systems with a rhombohedral stacking order
harbor nearly flat bands in their single-particle spectrum. We
propose ansatz states to describe the surface-localized states of
flat band electrons. The absence of kinetic dispersion near the
fermi level leaves the interaction as a dominate mechanism to govern
the low energy physics of a low density electron system. We build up
an effective lattice model in two interacting low-energy bands,
where the full terms of the Coulomb interaction, including those
long-range and off-diagonal parts, have been considered. The
interaction matrix coefficients in the many-body Hamiltonian model
are directly calculated for a trilayer system using orthonormal
Wannier basis. We then present a flat-band projection to yield an
interaction-only lattice model for flat band electrons. We find that
this limited model might energetically favor a ferromagnetic quantum
crystal under certain conditions.
\end{abstract}

\pacs{71.10.Fd, 71.27.+a, 73.21.Ac}

\maketitle

\section{Introduction}

Graphene based structures have drawn numerous attentions due to
their unique electronic
properties.\cite{novoselov:2004,castroneto:2009} The rapid technique
development enable people to engineer the graphene nanostructures in
special designs, yielding rich band structure features. In recent
years, great theoretical
\cite{guinea:2006,henrard06,min08,fzhang,mccann,yuan:2011,gelderen:2011,dora,peeters,koshino,kumar,otani:2010,chou,scherer,lang:2012,xu:2012,scherer:2012,cvetkovic:2012,dft,jung:2012,heikkila,liu:2012}
and experimental
\cite{exp01,exp02,exp03,exp04,exp05,exp06,exp07,tarucha,bao2010,norismatsu,mak,herrero,kumar:2011,lzhang,heinz11,heinz12,wbao,jhang}
interests have been focused on the graphene multilayer systems.
Different from the graphene monolayer, the band structure of the
multilayer graphene system depends on its stacking order, i.e., the
way to stacking the graphene sheets. Recently, the rhombohedral
stacking multilayer graphene has drawn intensive research interests
due to its intriguing band dispersion. It has two subbands near the
neutral system Fermi level, one conduction band and one valence band
with $|\epsilon| \sim k^N$ dispersion touching at $\epsilon=0$,
where $N$ is the layer number.\cite{guinea:2006} The rather flat
energy bands near $\epsilon = 0$ make the rhombohedral stacking
multilayer graphene susceptible to the interaction.\cite{fzhang}
Thus, the system  is instable towards quantum correlated phases,
such as superconductors or ferromagnets.\cite{heikkila,liu:2012}

Some recent experiments\cite{wbao,jhang} in rhombohedral stacking
graphene trilayer have shown the hints of a gapped ground state,
which is in sharp contrast with the gapless semiconducting ground
state suggested in noninteracting picture. Several symmetry-breaking
correlated states have been proposed as the candidates of the gapped
ground state, such as layered antiferromagnetic state, quantum
anomalous hall state, quantum spin hall state, and quantum valley
hall state.\cite{fzhang,scherer:2012,jung:2012} However, the
theoretical predictions strongly depend on the model and parameters
they chose. The detail properties of the ground state are still
under debate.

Flat band electrons of the rhombohedral stacking graphene system are of particular interest, since it is believed that the correlated ground state results from the interplay between the electron-electron interaction and the peculiar flat energy bands near the Fermi level. For a low density system the dispersion-less flat bands leave the Coulomb interaction predominantly rule the low energy physics. This calls for a comprehensive evaluation to the effects from all interaction terms, including those long-range density-density repulsion terms and leading off-diagonal terms, such as the direct spin exchange. The absence of the intra-band screening in a flat band suggests that these nonlocal interactions would be relevant. Studies have shown that these nonlocal interactions can lead to exotic correlated phases, such as quantum crystal and quantum liquids.\cite{wang:2011,wang:2012}
In this paper, we theoretically investigate the flat band electrons and their interaction in the rhombohedral stacking graphene multilayer system.  We  establish a set of many-body Hamiltonian models, which allow to appropriately include the effects from nonlocal interaction in addition to the Hubbard onsite term. Corresponding to the unique non-interacting band structure, a single-particle basis of Wannier functions is first constructed.  We then use our basis to
directly compute the matrix elements of a unscreened Coulomb interaction in two low-energy bands. A
projection protocol has been presented to approach an approximate interaction-only lattice model in the flat-band limit, which
are highly nontrivial, incorporating two bands, long-range
interactions, and spins. We argue that, at low densities, the
long-range part of the interaction in this limit model might support ferromagnetic
quantum crystals.

Our interaction model extends beyond the mean-field \cite{xu:2012} and renormalization group \cite{scherer:2012,cvetkovic:2012}studies, where a screened interaction with either the onsite Hubbard term or short-ranged interaction term is considered. Our study is also different from those with ${\it ab}$ ${\it initio}$ calculations \cite{dft} and Hartree-Fock approximations,\cite{jung:2012} which rely on certain local approximation to treat the nonlocal interaction and spin exchange terms. Alternatively, the Wannier basis allows us directly calculate these nonlocal terms.

This paper is organized as follows. In Section~\ref{singleparticle}
we consider the band structure that arises from the non-interacting
tight-binding model of rhombohedral
stacking graphene systems. An ansatz wave function has been proposed to
describe two flat bands.  In Section~\ref{wanniersection} we
construct localized single-particle basis states, orthonormal
Wannier functions, from carbon $\pi_{z}$ orbitals in graphene
honeycomb lattices. Section~\ref{bandmodel} uses the Wannier
functions to explicitly compute Coulomb interaction matrix elements
for two low-energy bands. Section~\ref{projection} defines a projection
scheme that limits the total many-body model to the flat-band
portion of the single-particle spectrum and discuss the possible low energy physics of this interaction-only lattice model. Section~\ref{summary} summarizes and looks forward to more accurate studies of the models constructed here.\\

\section{Flat Bands in rhombohedral stacking graphene sheets}
\label{singleparticle}

\begin{figure}[t]
\centerline{\includegraphics [width=3.5 in] {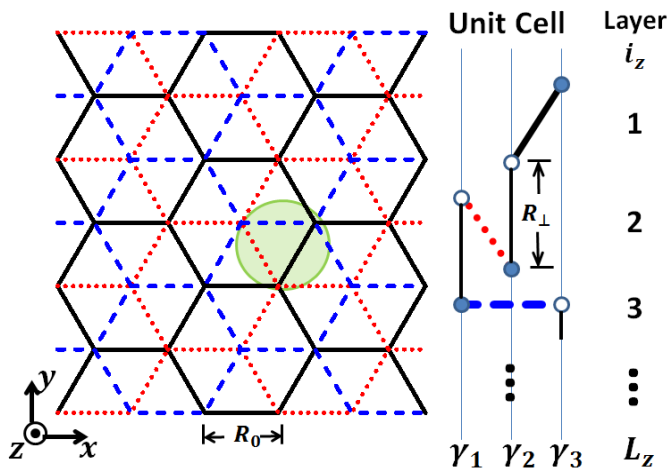}} \caption{(Color
online) Left: Schematic top-view of multiple layers of graphene
sheets in the rhombohedral stacking order. Lines in solid, dot and
dash types represent the in-plane carbon-carbon bonds at three
neighboring layers, counted from top to bottom in the $\hat{z}$
direction. The shaded area corresponds to a single unit cell. Right:
Schematic side-view of a unit cell in a triangular prism shape. The
solid and open circles stand for the atomic sites of the sublattice
$A$ and $B$, respectively. $\gamma_i$ are three corner axes of the
prism.} \label{lattice}
\end{figure}

We consider interacting electrons hopping among carbon sites of
rhombohedral graphene layers. In the left panel of
Fig.~\ref{lattice}, we schematically show the lattice of this
stacking system. Two neighboring graphene layers have a relative
in-plane shift along the carbon-carbon bond direction with the shift
distance equal to the bond length $R_0 \sim 1.42\mathrm{{\AA}}$.
After three successive shifts, the forth layer recovers the same
lattice as the first layer. We use $L_z$ to label the total number
of stacking layers and the layer separation is similar as the
graphite with $R_{\bot} \sim 3.35\mathrm{{\AA}}$. As shown in the
right panel of Fig.~\ref{lattice}, the primitive unit cell is in the
shape of a triangular prism with the total number of atom sites
$M=2L_z$. Each layer of the unit cell contains two sublattice sites
of $A$ and $B$ with perpendicular bonds to their counterpart
sublattice site at the neighboring layers. The array of unit cells
forms a two-dimensional Bravais triangular lattice with the lattice
length $R_c=\sqrt{3}R_0$.

In a simple non-interacting picture, the minimum single-particle tight-binding
Hamiltonian is given as: \cite{guinea:2006}
\begin{eqnarray}
H_{0}=-\sum_{\langle
n,m\rangle}(t_{mn}\hat{c}_{n}^{\dagger}\hat{c}_{m}^{\vphantom{\dagger}}+\text{h.c.}),
\label{H0}
\end{eqnarray}
where the sum is along carbon-carbon bonds and the hopping integrals
are taken $t_{\|}=3.16$ eV and $t_{\bot}=0.39$ eV for the intralayer
and interlayer hopping, respectively.\cite{dresselhaus:2002}  The
second-quantized operator $\hat{c}_{n}^{\dagger}$ creates a fermion
at a site $n$.  Labels $n$ and $m$ indicate lattice sites, in
contrast to labels for unit cells, $i,j,k,l$, used in the following.

\begin{figure}[t]
\centerline{\includegraphics [width=2.7 in] {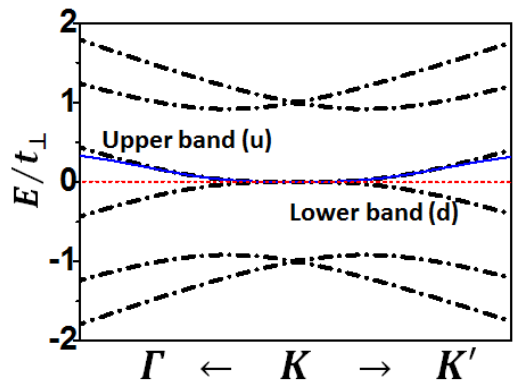}} \caption{(Color
online) The dot-dashed lines indicate the energy eigenvalues of
Eq.~(\ref{H0}) versus wavevector for the rhombohedral graphene
trilayer.  The solid line shows the approximate expression for the
energy, Eq.~(\ref{dispersion}).  Two flat bands form near the valley
points $K$ and $K'$.  In the large $L_{z}$ limit, the bands flatten.
} \label{band}
\end{figure}

Two bands near the Fermi level flatten around the corners
($\mathbf{K}$ and $\mathbf{K}'$ valley points) of the Brillouin zone
(BZ).  An example band structure for a trilayer system is shown in
Fig. 2. Crossing the Fermi level, the conduction band (upper band,
$u$) and valence band (lower band, $d$) are nearly degenerate with
in-plane wavevectors $\mathbf{q}$ (relative to the valley points) in
a region $|\mathbf{q}|<q^{\Delta}$ and form flat bands. For larger
number of stacking layers these bands can flatten considerably.

To model the two flat bands and examine the band width, we construct
analytical ansatz states in the linear combination of atom orbital
basis as $(\phi_{A},\phi_{B})^{T}$ with
$\phi_{A/B}=(\phi_{A/B,i_z=1}(\mathbf{q}),...,\phi_{A/B,i_z=L_z}(\mathbf{q}))$,
where the sites of sublattice $A$ ($B$) on the bottom (top) layer
have direct link to the neighboring layer. The indices $i_z$ marked
from 1 to $L_z$ represent the graphene layers from the topmost one
to the bottom as shown in Fig. 1.

For a wave function to be exact for $E=0$, the mathematical necessary
condition requires the wave function components between the neighboring layers to meet a certain relationship of
\begin{eqnarray}
\frac{\phi_{A,i_z}(\mathbf{q})}{\phi_{A,i_z+1}(\mathbf{q})}=\left(\frac{\phi_{B,i_z+1}(\mathbf{q})}{\phi_{B,i_z}(\mathbf{q})}\right)^*=p(\mathbf{q})^{-1},\nonumber\\
p(\mathbf{q})=-\frac{t_{\|}}{t_{\bot}}[e^{-iq_xR_0}+2\cos(\frac{\sqrt{3}}{2}q_y
R_0)e^{iq_xR_0/2}].
\end{eqnarray}
Note that at the valley points of the $\mathbf{K}$ and $\mathbf{K}'$, we have $|p(\mathbf{q})|=0$. The wave function is completely localized at two edge layers with the top layer occupied solely by the lattice $A$ and the bottom layer occupied solely by the lattice $B$. When the momentum is shifted away from the valley points, the wave function extends to the inner layers from the two edge layers. The ansatz wave functions in the vicinity of the valley points have the analytical form of $\Phi_{\pm}(\mathbf{q})=(\phi_{A},\pm \phi_{B})^{T}$ with $\phi_{A}=(1,p(\mathbf{q}),...,p(\mathbf{q})^{L_z-1})$ and $\phi_{B}=((p^{*}(\mathbf{q}))^{L_z-1},...,p^{*}(\mathbf{q}),1)$. In the general case with $|p(\mathbf{q}|\neq 1$, this ansatz state associates with a non-even occupation of the two sublattice sites on edge graphene layers.

Considering semi-infinite stacking layers of sublattices $A$ (edge
at the top surface) and $B$ (edge at the bottom surface), the
convergence of the wave function requires $|p(\mathbf{q})|<1$. This
determines the valid range of flat-band ansatz wave function with a
radius $q^{\Delta}/|\mathbf{K}|\approx(t_{\bot}/t_{\|})(\sqrt{3}/2\pi)$ in the limit of
$t_{\bot}/t_{\|}<1$. Here we see that an enhancement of the interlayer hopping
leads to a larger flat-band sector.

With the above ansatz states, the energy dispersion of bands
$\Gamma=u,d$ in the flat-band region can be computed explicitly:
\begin{eqnarray}
|E_{\Gamma}(\mathbf{q})|&\approx&\frac{|\Phi_{\pm}(\mathbf{q})^{T}H_{0}(\mathbf{q})\Phi_{\pm}(\mathbf{q})|}{|\Phi_{\pm}(\mathbf{q})|^2}\nonumber\\
&=&t_{\bot}\frac{|\text{Re}[p(\mathbf{q})^{L_z}]|(1-|p(\mathbf{q})|^2)}{1-|p(\mathbf{q})|^{2L_z}}
\label{dispersion}
\end{eqnarray}
with
\begin{equation}
H_{0}(\mathbf{q})=t_{\bot}\left(\begin{array}{cc}
0 & Q(\mathbf{q}) \\
Q^{\dagger}(\mathbf{q}) & 0
\end{array}\right),
\nonumber
\end{equation}
\begin{equation}
Q(\mathbf{q})=\left(\begin{array}{cccc}
-p^{*}(\mathbf{q}) & 0 & .. & 0 \\
1 & -p^{*}(\mathbf{q}) & 0 & .. \\
: & : & : & : \\
0 & .. & 1 & -p^{*}(\mathbf{q})
\end{array}\right).
\nonumber
\end{equation}
As shown in Fig. 2, the analytical dispersion Eq. (\ref{dispersion})
agrees with those calculated directly from the tight-bind
Hamiltonian in the vicinity of valley points, indicating the ansatz
wave function as an effective approximation to flat-band states.

With the equation (\ref{dispersion}), we can estimate the bandwidth
of the two nearly flat bands using the energy value at the flat-band
boundary $\mathbf{q}^{\Delta}$. In the large $L_z$ limit, the
bandwidth for states in the flat-band sector vanishes as:
\begin{eqnarray}
|E(\mathbf{q}\rightarrow \mathbf{q}^{\Delta})|\rightarrow
\frac{t_{\bot}}{L_z},
\label{bandwidth}
\end{eqnarray}
indicating that the band dispersion plays a small role with the
stacking number increasing. Such a vanishing bandwidth leaves the interaction as the dominant term in
the full many-body Hamiltonian of electrons.

For a dilute system with partially filled lattices, the lower-energy
physics of the electron system is mainly determined by the
single-particle basis states within the flat-band sectors near the
Fermi level. Thus, we project the Hamiltonian into the basis of
flat-band states in the approximation that $H_{0}$ adds an overall
constant energy shift to the spectrum. Our Hamiltonian model
becomes:
\begin{eqnarray}
H_{\text{total}}&=&\sum_{\mathbf{q}\in
\mathrm{BZ},\sigma,\Gamma}E_{\Gamma}(\mathbf{q})\hat{c}_{\mathbf{q}\sigma\Gamma}^{\dag}\hat{c}_{\mathbf{q}\sigma\Gamma}^{\vphantom{\dagger}}
+H_V\nonumber\\
&\rightarrow&\text{constant}+\mathcal{P}^{\dagger}_{\mathrm{FB}}H_V\mathcal{P}_{\mathrm{FB}}^{\vphantom{\dagger}},
\label{eqnprojection}
\end{eqnarray}
where the first equality is written in terms of the creation
(annihilation) operator $\hat{c}_{\mathbf{q}\sigma\Gamma}^{\dag}$
($\hat{c}_{\mathbf{q}\sigma\Gamma}^{\vphantom{\dagger}}$) for a
Bloch state at the wavevector $\mathbf{q}$ and band $\Gamma$, which
is related to the operator for a single-particle basis state in the
real space by the Fourier transform:
\begin{eqnarray}
\hat{c}_{j\sigma\Gamma}^{\dag}=\frac{1}{\sqrt{N}}\sum_{\mathbf{q}\in
\mathrm{BZ}}e^{i\mathbf{q}\cdot\mathbf{R}_j}\hat{c}_{\mathbf{q}\sigma\Gamma}^{\dag}.
\label{creator}
\end{eqnarray}
Here $\mathbf{R}_j$ is the lattice vector of the $j$-th unit cell,
$N$ denotes the number of unit cells in the system and $q$-space mesh in the BZ, and
$\sigma\in\{\uparrow, \downarrow\}$ labels spin.
$\mathcal{P}^{\dagger}_{\mathrm{FB}}$ denotes a projection into flat
bands such that the many-body eigenstates are constructed from Bloch
states in the flat-band sectors $|\mathbf{q}|<q^{\Delta}$.

To explore possible many-body ground states in the rhombohedral stacking
graphene system, we need construct an accurate form for
Eq.~(\ref{eqnprojection}) in the flat-band basis.  The
absence of dispersion excludes intra-band screening as in ordinary
Fermi liquids.\cite{dassarma:2007}  Thus, many-body eigenstates are
determined entirely by the interplay between various terms in the
interaction.  It is therefore crucial to accurately determine the
interaction terms in Eq.~(\ref{eqnprojection}) as prescribed by our
choice of single-particle basis.  In the next section, we describe
how to construct orthonormal Wannier functions to serve as
single-particle basis states.

\section{Single-Particle Basis States: Low-Energy Band Wannier Functions}
\label{wanniersection}

In this section we superpose overlapping carbon $\pi_z$ orbitals to form
orthogonal Wannier functions.  The Wannier functions will then be used to accurately determine interaction matrix
elements in later sections.

\begin{figure}[t]
\centerline{\includegraphics [width=3.4 in] {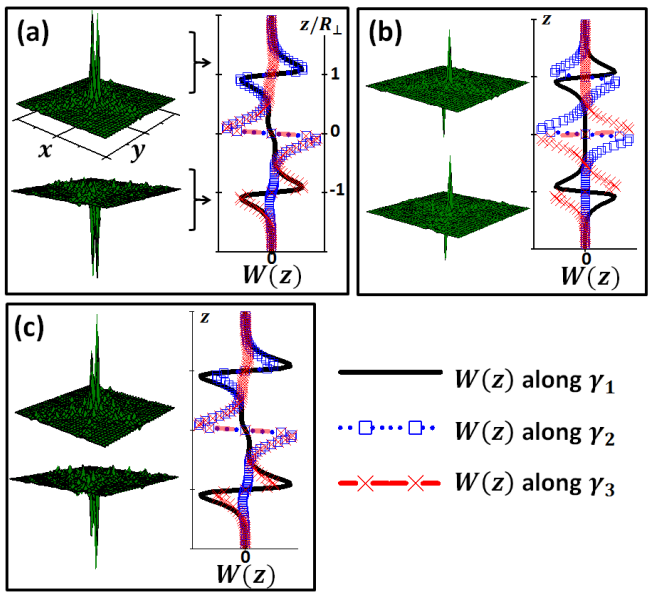}} \caption{(Color
online) Wannier functions of trilayer graphene sheets.  (a) $u$-band
case: Two three-dimensional plots on the left represent the
distribution of the Wannier functions in the $xy$ plane at the
$z$-positions right above the top layer and right below the bottom
layer, respectively. The twin peaks locate at two sublattice sites
of the edge layer in the original unit cell. The cartoon on the
right plots the distribution of the Wannier function in the
$\hat{z}$ direction along three corner axes of the original unit
cell. (b) The same plots as (a) but for $d$-band case. (c) $u$-band
Wannier function with adjusted parameters $t_{\bot}=t_{\|}$ to
emphasize the flat-band effect.}
 \label{wannier}
\end{figure}

In an isolated band the Wannier functions are given by
\begin{eqnarray}
 W_j(\mathbf{r})=W_0(\mathbf{r}-\mathbf{R}_j)=\frac{1}{N}\sum_{\mathbf{q}}e^{-i\mathbf{q}\cdot \mathbf{R}_j}\Psi_{\mathbf{q}}(\mathbf{r}),
\label{eq1}
\end{eqnarray}
where momenta $\mathbf{q}$ sum over $N$ discrete values inside the entire BZ. The Bloch functions are
$\Psi_{\mathbf{q}}(\mathbf{r})=\sum_{m=1}^{M}C_{m\mathbf{q}}\chi_{m\mathbf{q}}(\mathbf{r})$.

To make the contact with first principles calculations \cite{dft} we form Bloch functions from carbon
$\pi_z$ orbitals, $\phi(\mathbf{r})=\sqrt{\xi^5/\pi}ze^{-\xi r}$.
The basis states become
$\chi_{m\mathbf{q}}(\mathbf{r})=(1/\sqrt{N})\sum_{\mathbf{q}}e^{i\mathbf{q}\cdot\mathbf{R}_{j}}\phi(\mathbf{r}-\mathbf{r}_{mj})$,
where $\mathbf{r}_{mj}=\mathbf{R}_j+\mathbf{T}_m$ is the location of the $m$-th atom in the $j$-th unit
cell.

The coefficients $C_{m\mathbf{q}}$ and energy eigenvalues
$E(\mathbf{q})$ are obtained from diagonalization of the secular
equation:
\begin{eqnarray}
\left[ \tilde{O}^{-1} \tilde{H}(\mathbf{q}) \right] {\bf
C}_{\mathbf{q}}=E(\mathbf{q}){\bf C}_{\mathbf{q}},
\label{matrixequation}
\end{eqnarray}
where the matrix $\tilde{H}$ follows from the tight-binding
Hamiltonian $H_0$ with elements $\tilde{H}(\mathbf{q})_{mn}=\int
d\mathbf{r}\chi_{m\mathbf{q}}^{*}(\mathbf{r}) H_0
\chi_{n\mathbf{q}}(\mathbf{r})$. The overlap matrix $\tilde{O}$ are
taken as the the identity matrix in the tight-binding approximation.
The eigenvectors ${\bf
C}_{\mathbf{q}}\equiv\{C_{1\mathbf{q}},...,C_{M\mathbf{q}}\}^{T}$
yield the coefficients used in the definition of the Wannier
functions.

To construct orthonormal Wannier
function, a specific set of single-particle basis states are chosen
by enforcing $C_{m\mathbf{q}}$ at the edge atomic sites $m=1$ and
$m=M$ conjugate. The resulting Wannier functions $W_j(\mathbf{r})$
are \emph{real} and localized at $\mathbf{R}_j$ with the certain
symmetry between top and bottom portions in the stacking direction $\hat{z}$.

We can write the Wannier function at the origin as a summation over
all local atomic orbitals $\phi(\mathbf{r})$, i.e.,
\begin{eqnarray}
 W_0(\mathbf{r})=N_f \sum_{m=1}^{M}\sum_{i=0}^{N-1}\alpha_{mi}\phi(\mathbf{r}-\mathbf{r}_{mi})
\label{eq4}
\end{eqnarray}
with weights
$\alpha_{mj}=\sum_{\mathbf{q}}C_{m\mathbf{q}}e^{i\mathbf{q}\cdot\mathbf{R}_j}$
and the normalization constant $N_f$. A denser sampling in
momentum space (i.e., larger $N$) yields more accurate Wannier
functions. In practice, we find that the Wannier function has
already converged when taking $N=1261$ for $L_{z}=3$.

We can extend our construction of the Wannier functions to include
both the upper and lower bands. The Wannier functions of these two low-energy bands in a trilayer system  are shown in
Fig.~\ref{wannier}.  We note that these two Wannier functions mainly
localize at the original unit cell with the reflection symmetry
(antisymmetry) along a center line $(\sqrt{3},1,0)$ for the upper
(lower) band, decaying rapidly within several cell lengths. In the
plots of Wannier functions as a function of $z$-positions as shown
in Fig 3.~(a) and (b), Wannier functions mostly distribute in a
narrow region around each layer with the node on the layer. This is
due to the property of underlying $\pi_z$ orbitals. We note that
there exists a large portion of the Wannier function around the
middle layer, indicating the contribution from those extensive
states with momenta outside the flat-band region. Under the given
hopping parameters of $t_{\bot}/t_{\|} \sim 0.1$, two sublattices
near evenly occupy each layer.

Wannier functions built here integrate over the entire BZ. Thus, the
extensive states from the large non-flat region may shield the real
feature of the surface-localized state in the flat-band sectors.
Based on the analysis in the previous section, we have learned that
the size of the flat-band region and the flatness of the bands are
proportional to the hopping parameter $t_{\perp}$. To explore the
effect from the relevant flat-band states in the Wannier functions,
we study the case with the exaggerative parameter
$t_{\perp}=t_{\parallel}$. As shown in Fig. 3(c), the non-balanced
occupation between sublattices $A$ and $B$ at two surface layers
magnifies as the flat-band region expands, consistent with the
property of the ansatz flat-band state in the previous section.
Meanwhile, the relative portion of the extensive Wannier function
around the middle layer also reduces as expected. The flat-band
induced asymmetric occupation of two sublattices in the surface
layers may justify the origin of the gapped symmetry-breaking states
proposed by earlier theoretical
studies.\cite{fzhang,scherer:2012,jung:2012}

\section{Coulomb Interaction Model}
\label{bandmodel}

For a dilute system where the chemical potential lies between the
two nearly flat bands, the Coulomb interaction can in
principle favor occupancy of both bands or a single
band.  As a first approximation, we assume that the valence band is
inert and only the conduction band, $u$, is active.

An unscreened Coulomb interaction in a single band has a second-quantized many-body form of
\begin{eqnarray}
\sum_{i,j,k,l;\sigma,\sigma^{'}}\mathcal{V}_{ijkl}\hat{c}_{i\sigma}^\dag
\hat{c}_{j\sigma^{'}}^\dag
\hat{c}_{k\sigma^{'}}^{\vphantom{\dagger}}\hat{c}_{l\sigma}^{\vphantom{\dagger}},
\end{eqnarray}
where the operators $\hat{c}_{i\sigma}^{\dag}$
($\hat{c}_{i\sigma}^{\vphantom{\dagger}}$) create (annihilate) a
fermion with spin $\sigma$ in a Wannier state centered at the $i$-th
unit cell.  The matrix elements $\mathcal{V}$ are determined by the Wannier basis given in last section.
We can rewrite the above many-body Coulomb interaction in a
suggestive form:
\begin{eqnarray}
H_V^{u} &=&V_0\sum_{i} n_{i\uparrow}n_{i\downarrow}+\sum_{i<j}V_{ij}n_in_j-\sum_{i< j}J_{ij}\mathbf{S}_i\cdot\mathbf{S}_j\nonumber\\
&+&\frac{1}{2}\sum_{\{i,j\}\nsubseteq\{k,l\};\sigma,\sigma^{'}}V_{ijkl}\hat{c}_{i\sigma}^\dag
\hat{c}_{j\sigma^{'}}^\dag
\hat{c}_{k\sigma^{'}}^{\vphantom{\dagger}}\hat{c}_{l\sigma}^{\vphantom{\dagger}}.
\label{onebandH}
\end{eqnarray}
Here, the single-component and total density operators are
$n_{i\sigma}=\hat{c}_{i\sigma}^{\dag}\hat{c}_{i\sigma}^{\vphantom{\dagger}}$
and $n_i=n_{i\uparrow}+n_{i\downarrow}$, respectively. The spin
operators
$\mathbf{S}_i=(1/2)\sum_{\sigma\sigma^{'}}\hat{c}_{i\sigma}^{\dag}${\boldmath$\tilde{\sigma}$}$_{\sigma\sigma^{'}}\hat{c}_{i\sigma^{'}}^{\vphantom{\dagger}}$
are defined in terms of the Pauli matrices
{\boldmath$\tilde{\sigma}$}.

Eq.~(\ref{onebandH}) keeps all terms in the full Coulomb
interaction.  The first term is the ordinary onsite Hubbard term used in some mean-field studies of multilayer
graphene systems. \cite{lang:2012,xu:2012}  The second term captures the diagonal
portion of the Coulomb interaction at long range, which favors
certain charge order, such as a two-dimensional Wigner crystal.  The
absence of a dispersion in a low density system implies that this
term can be relevant and must be kept in accurate models.  The third
term, the direct exchange term, favors ferromagnetism for $J_{ij}
>0$.   The last term represents remaining off-diagonal terms due to
the Coulomb interaction, which are much small compared to the first
three terms for a single band according to our direct calculation.

The matrix elements in Eq.~(\ref{onebandH}) can be computed explicitly
using the Wannier basis in the $u$ band, as shown in the appendix,
Eqs.~(\ref{twobandcoefficients}). To calculate these integral equations, we have approximated the exponential part of the $\pi_z$ orbital as a
linear combination of three Gaussian functions:
$\phi(\mathbf{r})\approx
\sum_{s}\gamma_s(128\beta_s^5/\pi^3)^{1/4}ze^{-\beta_sr^2}$, where the parameters $\gamma_s$ and $\beta_s$ are  obtained from the STO-3G
package.\cite{emsl} Data for fitting the $\pi_z$ orbital with
$\xi=1.72$ are listed in Table \ref{tab1}. For numerical results
shown here and in the following tables, we use the Bohr radius,
$a_0=0.53 \mathrm{{\AA}}$, as the unit length and the Coulomb
energy $e^2/4\pi\epsilon a_0$ ($\sim$  27.2 eV in vacuum) as the
unit of energy.

\begin{table}[t]
\caption{Fitting parameters for the Gaussian approximation to the
$\pi_z$ orbital with $\xi=1.72$.} \centering
\begin{tabular}{|l|c|c|c|}
  \hline
  $s$            & 1          & 2          & 3 \\[-.1ex] \hline
  $\gamma_s$     & 0.15591627 & 0.60768372 & 0.39195739 \\[-.1ex]
  $\beta_s$      & 2.9412494  & 0.6834831  & 0.2222899  \\
  \hline
\end{tabular}
\label{tab1}
\end{table}

\begin{table}[t]
\caption{Matrix elements for one-band ($u$ band) case of the $L_z=3$
system with unit cell separations of up to $3R_{c}$.}
\centering
\begin{tabular}{|c|c|c|c|c|c|}
  \hline
  \multicolumn{6}{|l|}{$V_0$=3.587e-1} \\
  \hline
  $|\mathbf{R}_i-\mathbf{R}_j|/R_c$ & 1        &$\sqrt{3}$& 2        &$\sqrt{7}$& 3        \\
  $J_{ij}$                          & 2.136e-3 & 2.232e-3 & 9.703e-4 & 5.273e-4 & 6.075e-5 \\
  $V_{ij}$                          & 2.007e-1 & 1.462e-1 & 1.319e-1 & 1.080e-1 & 9.751e-2 \\
  \hline
\end{tabular}
\label{tab2}
\end{table}

Table~\ref{tab2} lists the coefficients computed for a trilayer
system.  As we see, all coefficients are positive and can be sorted
by $V_0 > V_{ij} > J_{ij} > 0$. These coefficients suggest that a
partially filled single band supports the formation of ferromagnetic
crystals.

However, the large Coulomb interaction may cause mixing between the
$u$ and $d$ bands.  We need consider a more comprehensive two-band
interaction model with Wannier functions in both the $u$ and $d$
bands.  The interaction Hamiltonian is dominated by the following terms:
\begin{eqnarray}
H_V^{ud}&=&\sum_{i,\Gamma}V_{0}^{\Gamma}n_{i\Gamma\uparrow}n_{i\Gamma\downarrow}+\sum_{i,\sigma,\Gamma \neq \Gamma^{'}}V_{0}^{'}n_{i\Gamma\sigma}n_{i\Gamma^{'}\sigma}\nonumber\\
&+&\sum_{i}\left (\sum_{\Gamma \neq \Gamma^{'}}V_{ii}^{'}n_{i\Gamma}n_{i\Gamma^{'}}-J_{ii}^{'}\mathbf{S}_{iu}\cdot\mathbf{S}_{id}\right )\nonumber \\
 &+&\sum_{i<j,\Gamma}(V_{ij}^{\Gamma}n_{i\Gamma}n_{j\Gamma}-J_{ij}^{\Gamma}\mathbf{S}_{i\Gamma}\cdot\mathbf{S}_{j\Gamma})\nonumber\\
 &+&\sum_{i<j}\sum_{\Gamma \neq \Gamma^{'}}(V_{ij}^{'}n_{i\Gamma}n_{j\Gamma^{'}}-J_{ij}^{'}\mathbf{S}_{i\Gamma}\cdot\mathbf{S}_{j\Gamma^{'}})\nonumber\\
 &+&\sum_{i<j}\sum_{\Gamma \neq \Gamma^{'}}\sum_{\sigma,\sigma^{'}}(
  V_{ij}^{''}\hat{c}_{i\Gamma\sigma}^{\dag}\hat{c}_{j\Gamma^{'}\sigma^{'}}^{\dag}\hat{c}_{j\Gamma\sigma^{'}}^{\vphantom{\dagger}}\hat{c}_{i\Gamma^{'}\sigma}^{\vphantom{\dagger}}\nonumber\\
&+&V_{ij}^{'''}\hat{c}_{i\Gamma\sigma}^{\dag}\hat{c}_{j\Gamma^{'}\sigma^{'}}^{\dag}\hat{c}_{i\Gamma^{'}\sigma^{'}}^{\vphantom{\dagger}}\hat{c}_{j\Gamma\sigma}^{\vphantom{\dagger}}).
\label{fullmodel}
\end{eqnarray}
We have checked, by direct calculations, that other terms involving
three or four centers are much smaller than terms kept here. In Eq. \ref{fullmodel}, we see the Hubbard and ferromagnetic terms
as in the one-band analysis.  Besides, we have the
non-trivial band exchange as the last term. The integral equations for all coefficients
in Eq. \ref{fullmodel} are listed in the Appendix.

\begin{table}[t]
\caption{Same as the Table~\ref{tab2} but for the two-band case.}
\centering
\begin{tabular}{|c|c|c|c|}
  \hline
  \multicolumn{2}{|l}{$V^{d}_{0}$=1.495e-1}  & \multicolumn{2}{l|}{$V^{u}_{0}$=3.587e-1}\\
  \multicolumn{2}{|l}{$V^{'}_{ii}$=2.192e-1} & \multicolumn{2}{l|}{$V^{'}_{0}$=4.932e-2}\\
  \multicolumn{4}{|l|}{$J_{ii}^{'}$=9.864e-2}\\
  \hline
  $|\mathbf{R}_i-\mathbf{R}_j|/R_c$ & 1 & 2 & 3  \\
  $V_{ij}^{d}$   & 8.419e-2 & 5.527e-2 & 4.079e-2 \\
  $V_{ij}^{u}$   & 2.007e-1 & 1.319e-1 & 9.751e-2 \\
  $V_{ij}^{'}$   & 1.304e-1 & 8.562e-2 & 6.313e-2 \\
  $J_{ij}^{d}$   & 8.877e-4 & 3.965e-4 & 2.443e-5 \\
  $J_{ij}^{u}$   & 2.136e-3 & 9.703e-4 & 6.075e-5 \\
  $J_{ij}^{'}$   & 2.726e-4 & 2.052e-4 & 2.058e-5 \\
  $V_{ij}^{''}$  & 7.467e-4 & 7.920e-5 & 2.994e-5\\
  $V_{ij}^{'''}$ & 6.885e-4 & 3.100e-4 & 3.853e-5\\
  \hline
\end{tabular}
\label{tablelz3}
\end{table}

Table~\ref{tablelz3} shows numerically computed coefficients for the
two-band model Eq.~\ref{fullmodel} in a trilayer system. Rows 1-3
exhibit several leading terms of the diagonal components of Coulomb
interaction, which primarily determine the charge degrees of
freedom. Rows 4-6 govern the spin degrees of freedom.  The positive
elements support ferromagnetism.  The last two rows give rise to
band exchange effects.

The calculated coefficients of the onsite Coulomb repulsion have
values of $2 - 5$ eV with an estimated effective dielectric constant
$\epsilon=2$ in graphene systems, which are consistent with the
parameter range in a mean-field analysis \cite{xu:2012} for the
experimentally observed energy gap.\cite{wbao,jhang} We also note
that the long-range interaction terms of up to the fifth nearest
neighbors (rows 1-3) have a magnitude comparable to the onsite
terms, indicating the effective interaction range could be much
longer than the usual screened interaction treatments with up to
nearest or next-to-nearest neighbors. Based on the energetic
argument these long-range terms are relevant and should be included
to discuss the possible low energy states of a dilute system.

\section{Flat-Band Projection}
\label{projection} In this section we construct a set of operators
that allow flat-band projection of the many-body  Hamiltonian model
constructed in the previous sections. We then discuss the possible
low energy physics under certain conditions.

To enforce  flat-band projection we limit all $q$-space sums to the
flat-band region (FBR) $|\mathbf{q}|<q^{\Delta}$.  We first consider
a state operator in a single band that limits itself to the FBR:
\begin{eqnarray}
\hat{b}_{j\sigma}^{\dag}\equiv\frac{1}{N}\sum_{l}\sum_{\mathbf{q}\in
\mathrm{FBR}}e^{i\mathbf{q}\cdot(\mathbf{R}_j-\mathbf{R}_l)}\hat{c}_{l\sigma}^{\dag}.
\label{proj-creator}
\end{eqnarray}
This operator creates states centered around the unit cell at
$\mathbf{R}_j$ while can overlap considerably with neighbors, indicating that
the projection into a flat band delocalizes basis states.  In
the limit that the flat band encompasses the entire Brillouin zone, we have
$\hat{b}_{j\sigma}^{\dag}\rightarrow\hat{c}_{j\sigma}^{\dag}$.

We can then rewrite our model in terms of projected density and spin
operators.  The single-component and total projected density
operators are $\rho_{i\sigma}\equiv
\hat{b}_{i\sigma}^{\dag}\hat{b}_{i\sigma}^{\vphantom{\dagger}}$ and
$\rho_i\equiv\rho_{i\uparrow}+\rho_{i\downarrow}$, respectively. The
projected spin operators are defined as:
\begin{eqnarray}
\slashed {\mathbf S }_{j} \equiv
\frac{1}{2N}\sum_{\sigma\sigma^{'}}\sum_{\mathbf{q},\mathbf{q'}\in
\mathrm{FBR}}e^{i(\mathbf{q}-\mathbf{q'})\cdot\mathbf{R}_j}\hat{c}_{\mathbf{q}\sigma}^{\dag}{\boldsymbol
{\tilde{\sigma}}}_{\sigma\sigma^{'}}\hat{c}_{\mathbf{q'}\sigma^{'}}^{\vphantom{\dagger}}.
\label{proj-spin}
\end{eqnarray}
Note that these projected operators do not exhibit
ordinary commutation relations because the underlying basis states
are delocalized.

The  projected Hamiltonian can be rewritten entirely in terms of the
above projected operators. Starting from an unprojected interaction model, we
impose projection using the following replacements: $c\rightarrow
b,n\rightarrow\rho,$ and ${\mathbf S}\rightarrow\slashed {\mathbf
S}$. Considering the intrinsic energetic ordering as shown in the table ~\ref{tablelz3},  we rewrite the two-band interaction Hamiltonian in the
projected space:
\begin{eqnarray}
\mathcal{P}^{\dagger}_{\mathrm{FB}}H_V^{ud}\mathcal{P}^{\vphantom{\dagger}}_{\mathrm{FB}}&=&\sum_{i,\Gamma}V_{0}^{\Gamma}\rho_{i\Gamma\uparrow}\rho_{i\Gamma\downarrow} +\sum_{i,\sigma,\Gamma \neq \Gamma^{'}}V_{0}^{'}\rho_{i\Gamma\sigma}\rho_{i\Gamma^{'}\sigma}\nonumber \\
 &+&\sum_{i, j,\Gamma,\Gamma'}\left(\overline{V}_{ij}^{\Gamma,\Gamma'}\rho_{i\Gamma}\rho_{j\Gamma'}-\overline{J}_{ij}^{\Gamma,\Gamma'}\slashed {\mathbf S}_{i\Gamma}\cdot\slashed {\mathbf S}_{j\Gamma'}\right)\nonumber\\
 &+&H_{\text{Band-exch}},
\label{Vsimplified}
\end{eqnarray}
where we  have redefined the diagonal Coulomb terms:
$\overline{V}_{i<j}^{\Gamma\neq\Gamma'}\equiv V'_{ij}$,
$\overline{V}_{ii}^{\Gamma=d,\Gamma'=u}\equiv V'_{ii}$, and
$\overline{V}_{i<j}^{\Gamma=\Gamma'}\equiv V^{\Gamma}_{ij}$,
otherwise $\overline{V}_{ij}^{\Gamma,\Gamma'}=0$. We have also
redefined the off-diagonal exchange terms:
$\overline{J}_{i<j}^{\Gamma\neq\Gamma'}\equiv J'_{ij}$,
$\overline{J}_{ii}^{\Gamma=d,\Gamma'=u}\equiv J'_{ii}$, and
$\overline{J}_{i<j}^{\Gamma=\Gamma'}\equiv J^{\Gamma}_{ij}$,
otherwise $\overline{J}_{ij}^{\Gamma,\Gamma'}=0$.  The last term in
Eq.~(\ref{Vsimplified}) corresponds to the last term in
Eq.~(\ref{fullmodel}).

Let us consider a conditional dilute system with the Fermi enery
away from the charge neutrality point, where the band far away from
Fermi level is approximately inert and the band-exchange terms can
be ignored. The first three terms in Eq.~(\ref{Vsimplified}) will
impose a rigid charge order analogy to the classical Wigner crystal.
However, here the charges are significantly delocalized.  A finite
overlap among neighbors implies that the charges exist in a
superposition of several different unit cells at once, indicating a
quantum crystal. The forth term corresponds to an effective
Heisenberg model which favors ferromagnetism between neighboring
cell spins. Thus, the projected system favors the ground state as
ferromagnetic quantum crystals. Correspondingly, low energy spin
excitations could be ferromagnetic magnons.\cite{wang:2012} At low temperature, the ferromagnetic order among two-dimensional cell spins could be detectable using the magnetic scanning probe microscopy technique, such as magnetic force microscopy and spin-polarized scanning tunneling microscopy.\cite{wiesendanger} We note
that this in-plane ferromagnetic order is also suggested by several
theoretical
models~\cite{xu:2012,scherer:2012,cvetkovic:2012,dft,jung:2012} in
the distinct system where the Fermi level is near the charge
neutrality point and the correlation between both flat bands
involves.

\section{Summary and Outlook}
\label{summary}

We construct interaction lattice model for flat band electrons in
rhombohedral stacking graphene layers.  An ansatz wave function has
been proposed to describe the properties of flat-band states
emerging in the single-particle spectrum of the system.  A
single-particle basis of orthonormal Wannier functions was built
from carbon $\pi_{z}$ orbitals of the underlying graphene honeycomb
lattice. We use this single-particle basis to explicitly compute the
Coulomb matrix elements. The total model, Eqs.~(\ref{fullmodel}),
was then projected into the flat bands, suggesting a ferromagnetic
quantum crystal ground state under certain assumptions.

Numerical results are shown here for the trilayer system. However,
the formulae of our model and approach are general to the
rhombohedral graphene multilayer system. In a separate calculation
with more layers, we have found the similar feature of the Wannier
basis and the relative order among interaction coefficients. Our
flat-band model, Eq.~(\ref{Vsimplified}), sets the stage for more
accurate study with a combination  of variational studies and
diagonalization to verify possible ground and excited
states.\cite{wang:2011}

We also want to stress the difference between the work presented
here and a previous mean-field study.\cite{xu:2012} Our interaction
model includes a full consideration of the nonlocal interaction
terms from two low-energy bands. The mean-field study \cite{xu:2012}
takes interaction contribution from all bands but only counts the
onsite interaction term. Our model can be applicable in the limit
case with the large stacking number and weak interaction. There the
low-energy properties of the system are most relevant to two
extremely flat band portions. In an otherwise case where the
interaction is strong and the screening effect from those dispersive
higher-energy bands are not negligible, the mean-field treatment
would be justified.

The constructed model focuses on key physics of interacting flat
bands but excludes a number of realistic effects. For example, longer-range hopping can cause the trigonal warping effect and other effects, which distort the flat bands. The experimental conditions, such as
defects and substrate disorder can also destroy the flat-band
approximation. We apply the flat-band limit considering that these deformations are less than the estimated bandwidth in Eq.~(\ref{bandwidth}). In addition, inter-band screening from nearby bands could lead to corrections to the Coulomb interaction studied here.

\section{Acknowledgements}

We acknowledge part of financial support from HKSAR RGC grant HKU
701010. JHG is supported by National Natural Science Foundation of China (Project No. 11274129).

\section{Appendix}
The coefficients in Eqs.~\ref{onebandH} and \ref{fullmodel} are
given by:
\begin{eqnarray}
V_{0}^{\Gamma}&=&\int\frac{d^3\mathbf{r}d^3\mathbf{r'}}{|\mathbf{r}-\mathbf{r'}|}|W_{0\Gamma}(\mathbf{r})W_{0\Gamma}(\mathbf{r'})|^2, \nonumber\\
J_{ij}^{\Gamma}&=&2\int\frac{d^3\mathbf{r}d^3\mathbf{r'}}{|\mathbf{r}-\mathbf{r'}|}W_{i\Gamma}^{*}(\mathbf{r})W_{j\Gamma}(\mathbf{r})W_{i\Gamma}(\mathbf{r'})W_{j\Gamma}^{*}(\mathbf{r'}),  \nonumber\\
V_{ij}^{\Gamma}&=&\int\frac{d^3\mathbf{r}d^3\mathbf{r'}}{|\mathbf{r}-\mathbf{r'}|}|W_{i\Gamma}(\mathbf{r})W_{j\Gamma}(\mathbf{r'})|^2-\frac{1}{4}J_{ij}^{\Gamma},  \nonumber\\
J_{ij}^{'}&=&2\int\frac{d^3\mathbf{r}d^3\mathbf{r'}}{|\mathbf{r}-\mathbf{r'}|}W_{iu}^{*}(\mathbf{r})W_{jd}(\mathbf{r})W_{iu}(\mathbf{r'})W_{jd}^{*}(\mathbf{r'}),  \nonumber\\
V_{ij}^{'}&=&\int\frac{d^3\mathbf{r}d^3\mathbf{r'}}{|\mathbf{r}-\mathbf{r'}|}|W_{iu}(\mathbf{r})W_{jd}(\mathbf{r'})|^2-\frac{1}{4}J_{ij}^{'},  \nonumber\\
V_{0}^{'}&=&\frac{1}{2}J_{ii}^{'},  \nonumber\\
V_{ij}^{''}&=&\int\frac{d^3\mathbf{r}d^3\mathbf{r'}}{|\mathbf{r}-\mathbf{r'}|}W_{iu}^{*}(\mathbf{r})W_{id}(\mathbf{r})W_{ju}(\mathbf{r'})W_{jd}^{*}(\mathbf{r'}),  \nonumber\\
V_{ij}^{'''}&=&\int\frac{d^3\mathbf{r}d^3\mathbf{r'}}{|\mathbf{r}-\mathbf{r'}|}W_{iu}^{*}(\mathbf{r})W_{ju}(\mathbf{r})W_{id}(\mathbf{r'})W_{jd}^{*}(\mathbf{r'}),  \nonumber\\
V_{ijkl}&=&\int\frac{d^3\mathbf{r}d^3\mathbf{r'}}{|\mathbf{r}-\mathbf{r'}|}W_{iu}^{*}(\mathbf{r})W_{lu}(\mathbf{r})W_{ju}^{*}(\mathbf{r'})W_{ku}(\mathbf{r'}).
\label{twobandcoefficients}
\end{eqnarray}
The last term is used only in Eq.~(\ref{onebandH}).

\end{document}